# A possible scenario for the mechanism of high-$T_c$ superconductivity based on experimental data

A. Mourachkine[*]

*University of Cambridge, J. J. Thomson Ave. Cambridge, CB3 0HE, UK*

**Abstract**

The issue of the mechanism of high-$T_c$ superconductivity remains open. In this contribution, we propose a new scenario for the mechanism of superconductivity in cuprates based on analysis of experimental data, mainly tunneling, neutron scattering and muon-spin-relaxation data, made earlier (see e.g. Mod. Phys. Lett. B 19 (2005) 743). A specific feature of this scenario is the mechanism of the establishment of long-range phase coherence among Cooper pairs, based on recent experimental data obtained in nonsuperconducting materials.

In spite of tremendous number of accumulated experimental data, at present, 20 years after the discovery of high-$T_c$ superconductivity (HTSC) [1], there is no generally accepted theoretical model describing this remarkable phenomenon. Soon after the discovery of HTSC, it became evident that there is a clear microscopic difference between the BCS-type superconductivity (SC) and HTSC, namely, that they have a different origin and that different criteria are required for HTSC than for conventional SC. As for the bulk characteristics, the controlling factor in cuprates seems only to be the hole density in $CuO_2$ planes. One of the main problems in selecting the right model for HTSC is that, for every existing *theoretical* model of HTSC, one can always find a set of experimental data which is not in agreement with the model. Such a situation should always be the case if HTSC is not just one-step phenomenon but consists of, at least, two phenomena "peacefully" co-existing. This is the main idea of this contribution, and here we discuss a **sketch** of a possible scenario for the mechanism of HTSC.

Analysis of experimental data puts restrictions on the future model, narrowing the choice for theorists. Let us start with analysis of experimental data, made earlier in a recent review, a book and a few papers [2-6]. In fact, we need *just* two conclusions obtained for SC in cuprates [2-6]. Analysis of tunneling, neutron scattering (INS) and muon-spin-relaxation data shows that spin fluctuations mediate the long-range phase coherence in cuprates [2-5]. At the same time, analysis of tunneling, INS, X-ray scattering and angle-resolved photoemission (ARPES) data indicates that the Cooper pairs in cuprates are most likely topological excitations, and phonons *seem* to be responsible for the pairng [2-4,6]. (For more details, see the references [2-6].)

Before continuing, it is worth to remind that SC requires the electron pairing and the establishment of long-range phase coherence among the pairs [3]. Moreover, in any SCor, the mechanism of pairing and the mechanism of long-range phase coherence are different [3]. For example, in conventional SCs, phonons are responsible for the pairing, while the overlap of pair wavefunctions (the Josephson coupling) mediates the phase coherence.

Turning our attention back to the aforementioned analysis of experimental data obtained in cuprates, one can immediately underline that it is not easy "to reconcile" the two conclusions. Nevertheless, *it is possible* if we assume that there are **two**, more or less independent, processes in cuprates. One of them is the formation of *incoherent* electron (hole) pairs, and the second process is the Bose-Einstein condensation (BEC) of magnetic excitations, for example, magnons. The BEC of magnons was recently observed in a number of antiferromagnetic compounds [7-11]. If this is the case for cuprates, and the Cooper pairs are coupled to magnetic excitations, then the pairs can adjust their phases, i.e. establish the phase coherence, through the long-range phase coherence of the magnon-BEC condensate. The Cooper pairs can be coupled to magnetic excitations through the amplitude of their wavefunctions, or only through the phase, or through both.

---
[*] *E-mail:* andrei_mourachkine@yahoo.co.uk

Why should we make a link between HTSC and magnon BEC? Because a few characteristics of the two phenomena are very similar. Consider them briefly. (i) The magnon BEC is characterized by

$$B_c \sim T_c^{3/2}(B_c), \qquad (1)$$

where $T_c$ is the critical temperature and $B_c$ is the critical magnetic field [7-11]. If in conventional SCs, $B_c \sim T_c^2$, in cuprates, the dependence is rather $B_c \sim T_c^{\sqrt{2}}$, obtained in low-$T_c$ cuprates [3]. Thus, as one can see the two dependences are similar (such a power dependence in cuprates can also be explained by the presence of defects or impurities). (ii) For most of the known antiferromagnets, the values of $B_c(0)$ can be extremely high, above 100 T. The values of $B_{c2}(0)$ in hole-doped cuprates are incredibly high too [3]. (iii) The magnon BEC is accompanied by the appearance of a collective oscillation (a Goldstone mode) of the BEC condensate with a spin of 1 [7-11]. In cuprates, below $T_c$ there is a so-called magnetic resonance peak which is a magnetic excitation with a spin of 1; however, it propagates at some finite energies of the order of tens of m$e$V (depending on hole concentration) [3] (normally, a Goldstone mode propagates at the ground state). (iv) At the transition temperature, the magnon BEC is accompanied by a jump in the specific heat, as shown in Fig. 1 for $Cs_2CuCl_4$. The curves in Fig. 1 look very similar to those obtained in cuprates (see Fig. 2 in Ref. [12]). Secondly, the $C_{mag}(T)$ dependences in Fig. 1 are shown exclusively close to the critical field. As one can see in Fig. 1, the value of the jump at $T_c$ increases as $B$ decreases; so, the jump value will be even larger at lower fields. It is well known that, in cuprates, the value of the jump in specific heat is larger than that in conventional SCs [12,3]. (v) In a different class of magnetic compounds, spin-gap antiferromagnetic dimers, such as $TlCuCl_3$ [8,9,11] and $BaCuSi_2O_6$ [10], the magnon-BEC phase has a shape of a dome [10,11] which is similar to the SC dome of cuprates (however, in the dimers, the dome is field-induced). It is worth to remind that cuprates have a spin gap [13], and in the striped phase, $Cu^{2+}$ spins which are located between charge stripes form antiferromagnetically ordered dimers [13], as shown in Fig. 2. So, taking all these facts together, one can conclude that it is worth to consider such a scenario for cuprates.

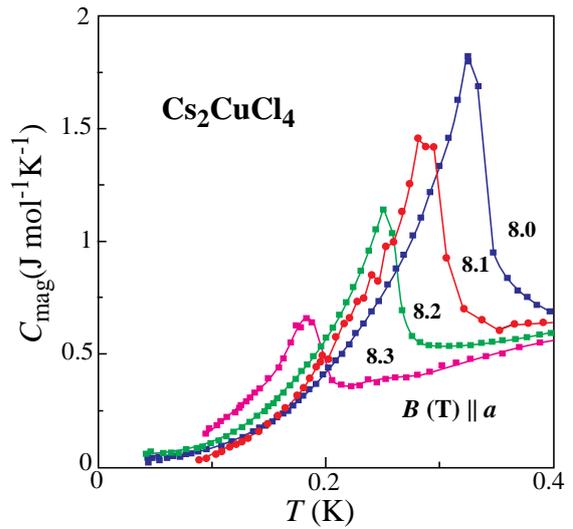

Fig. 1. Magnetic specific heat of $Cs_2CuCl_4$ [7].

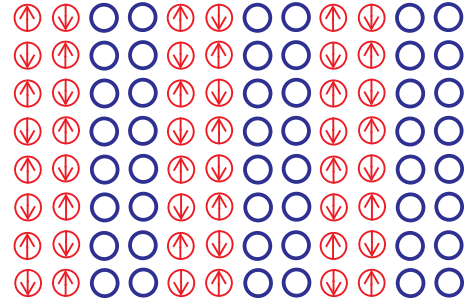

Fig. 2. Cartoon of stripe order in $CuO_2$ planes. Only Cu sites are shown, with arrows indicating ordered magnetic moments and circles indicating hole-rich charge stripes [13].